\documentclass[twocolumn,prl,showpacs]{revtex4}
\usepackage{graphicx}

\usepackage{amssymb}
\usepackage{amsmath}
\usepackage[latin1]{inputenc}
\usepackage{bm}

\begin{document}

\title{Counting statistics for mesoscopic conductors with internal degrees of freedom}
\author{Christopher Birchall  and Henning Schomerus}
\affiliation{Department of Physics, Lancaster University,
Lancaster,  LA1 4YB, United Kingdom}
\date{\today}
\begin{abstract}
We consider the transport  of electrons passing through a
mesoscopic device possessing internal dynamical quantum degrees of
freedom. The mutual interaction between the system and the conduction
electrons contributes to the current fluctuations, which we describe in terms of
full counting statistics. We identify conditions where this discriminates coherent from incoherent internal dynamics,
and also
identify and illustrate
conditions under which the device acts to dynamically bunch
transmitted or reflected electrons, thereby generating  super-Poissonian noise.
\end{abstract}

\maketitle

The probabilistic nature of quantum transport constitutes a fundamental
source of current fluctuations, whose study provides
detailed information about the carrier dynamics \cite{blanter:buttiker}.
 A comprehensive characterization of these
fluctuations is provided by the framework of full counting statistics
(FCS), which delivers a unifying perspective on the conductance $G=I/V$ (where $I$ is the current and
$V$ is the bias voltage), the shot noise $P=2 e I  F $ (where $F$ is the Fano
factor, with $F=1$ signifying the Poissonian statistics of uncorrelated
carriers), and higher-order stationary current correlations \cite{nazarov:blanter}. In the regime of single-particle
electronic transport close to equilibrium, the Pauli principle fundamentally
constrains  the current fluctuations, which results in sub-Poissonian noise with Fano factor $F<1$. Repulsive
interactions between the charge carriers generally result in a further
reduction of the noise \cite{bagrets:nazarov}. On the other hand,
super-Poissonian statistics can be achieved in strongly correlated systems,
where interactions enable the formation of a charge density wave \cite{bagretsreview}.
The noise
can also be enhanced via external means, e.g., by the presence of classical
multi-state fluctuators \cite{hassler:lesovik:blatter}.

\begin{figure}[t]
\includegraphics[width=\columnwidth]{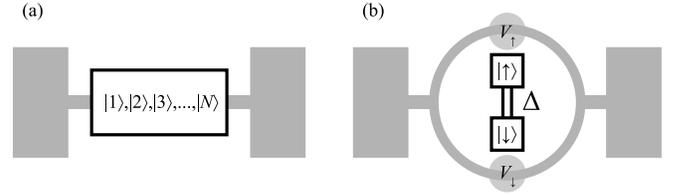}
\caption{\label{fig:1} (a) Sketch of a mesoscopic device with internal degrees of freedom, attached to leads
contacted by electronic reservoirs.
(b) Realization of such a system in terms of a coherent which-path interferometer,
consisting of an excess electron in a double quantum dot (states $|\uparrow\rangle$, $|\downarrow\rangle$, tunnel splitting $\Delta$), which interacts via a state-dependent
potential $V_{\uparrow,\downarrow}$ with
charge carriers passing through an Aharonov-Bohm ring.
}
\end{figure}

\begin{figure}[top]
\includegraphics[width=\columnwidth]{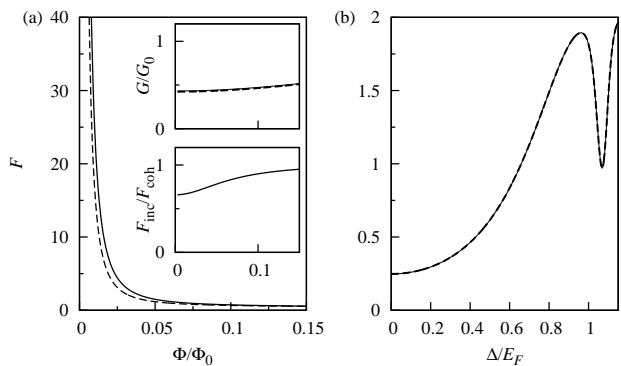}
\caption{\label{fig:results}  Fano factor for the coherent which-path interferometer sketched in Fig.\ \ref{fig:1}(b), where
(a) shows results as function of the magnetic flux $\Phi$ penetrating the Aharonov-Bohm ring, for fixed tunnel splitting $\Delta=0.01\,E_F$,
and (b) shows results as function of $\Delta$, for fixed $\Phi=h/2e$. Solid (dashed) lines: coherent (incoherent) internal dynamics. Insets in (a): dimensionless conductance (top), ratio of incoherent and coherent Fano factors (bottom).
The result are obtained by propagation  of minimal-uncertainty wave packets with energy spread $\Delta E=0.007\, E_F$.
The interaction $V_{\uparrow,\downarrow}$ is modeled by an effectively impenetrable barrier of height $20 E_F$, which blocks one of the two arms (of length $L$,
with $\hbar^2/2mL^2=0.061 E_F$) depending on the occupation of the double dot.
}
\end{figure}

Many quantum conductors of present interest, such as  molecules with different conformations and
nanoelectromechanical systems, possess internal quantum degrees of freedom which can be associated to  multiple internal states of a mesoscopic device [see Fig.\ \ref{fig:1}(a)].
These degrees of freedom are not directly associated to the charge carriers,
but cannot be considered external because of the backaction of the passing
carriers.
The purpose of this work is to demonstrate that these internal dynamics can be effectively probed via their influence on the counting statistics.
We find that even though the internal degrees of freedom do not introduce
strong correlations in the conventional sense \cite{koch}, they can result in
super-Poissonian statistics with $F\gg 1$. The transport then takes the form of dynamically
created trains of transmitted or reflected electrons. We
identify conditions of ideal dynamical bunching and show that these can
be realized in a model system,  a coherent which-path interferometer \cite{schomerus:noat,schomerus:robinson} consisting of an Aharonov-Bohm ring which is
electrostatically coupled to an excess electron in a double quantum dot [see Fig.\ \ref{fig:1}(b), and results in Fig.\ \ref{fig:results}].
We also identify a regime where the transmitted bunches consist of exactly two electrons, with $F=2$.

These phenomena are identified  by extension of previous works on
counting statistics in passive systems, in particular, counting statistics of
conductors coupled to quantum detectors (electrometers) \cite{averin:sukhorukov,brandes},  and the wave-packet approach \cite{hassler2008},
which provides a convenient dynamical perspective on FCS.
The resulting expression for FCS takes the form of a
generalized propagation of the density matrix for the internal degrees of  freedom.
A key conceptual ingredient appearing in our expressions is the extent of
coherence maintained throughout the evolution of the internal system
dynamics, which manifests itself in electronic interference
contributions associated to different dynamical histories of the internal
system state.

\emph{Model and framework.---}
We consider electronic transport through a mesoscopic system with
internal degrees of freedom, as  schematically depicted  in Fig.\
\ref{fig:1}(a). The system possesses  $N$ internal states
$|n\rangle$, $n=1,2,3,\ldots,N$, which affect the transport of electrons
passing through the system. In terms of creation and annihilation operators
$\psi^\dagger$, $\psi$  of electrons in momentum (${\bf k}$) and position
(${\bf r}$) representation, a reference Hamiltonian representative of this
problem takes the form
\begin{eqnarray}H&=&\int d{\bf k} \epsilon({\bf k})\psi^\dagger({\bf k})\psi({\bf k})+\sum_{nm} H^{\rm sys}_{nm}|n\rangle \langle m|\nonumber\\
&&+\sum_n\int d{\bf r}\psi^\dagger({\bf r})\psi({\bf r}) V_n({\bf r})|n\rangle\langle n|.
\end{eqnarray}
Here, the first term describes the kinetic energy of  electrons with
dispersion relation $\epsilon({\bf k})$, which is independent of the internal
system state. The second term is the Hamiltonian of the isolated internal
system degrees of freedom. The last term  describes the interaction between
these degrees of freedom and the electrons, which here is formulated in terms
of a  potential energy $V_n$ whose state-dependent part is assumed to be
confined to the scattering region (the potentials $V_n$ may also contain a
state-independent part outside this region, which can model, e.g., some
features of the leads).

Note that we do not assume that the isolated system Hamiltonian commutes with
the interaction part. Therefore, even when the internal degrees of freedom
are initially prepared in an eigenstate $|\sigma\rangle$ of $H^{\rm sys}$
(corresponding to energy ${\cal E}_\sigma$, where $\sigma=1,2,3,\ldots,N$),
the interaction with scattered electrons will induce dynamics
into the internal system. Moreover, since the system is not driven
externally, total energy conservation implies that any change of internal
system energy will be compensated by a change of energy of the transmitted
or reflected electrons, which amounts to inelastic (but still coherent)
scattering. We are interested in the consequence of these phenomena for the
average current and its temporal fluctuations.

A general framework for the statistical description of the electronic
transport is provided by full counting statistics,
which delivers information about the probability $P(Q;{\cal T})$ that a
number $Q$ of  charge carriers are transmitted during a time interval ${\cal
T}$. This information is encoded into the moment generating function
\begin{equation}
\chi(\lambda)=\sum_{Q=0}^\infty P(Q;{\cal T})\exp(\lambda Q)=\sum_{k=0}^\infty\langle Q^k\rangle\frac{\lambda^k}{k!},
\label{generatingfunctiondefinition}
\end{equation}
where $\langle Q^k\rangle$ denotes the $k$th moment of $Q$. Of particular interest are the associated cumulants
of order $m$,
\begin{equation}
\langle\langle Q^m \rangle\rangle=\left.\frac{\partial^m}{\partial\lambda^m}\ln \chi(\lambda)\right|_{\lambda=0},
\end{equation}
which for large times increase linearly in ${\cal T}$, $
\langle\langle Q^m \rangle\rangle\sim ({\cal T}/\tau)q_m$.
Here $1/\tau$ is the attempt frequency of transmission for incoming electrons ($\tau=h/eV$ for bias voltage $V$);
the coefficient $q_1=g=G/G_0$ then delivers the dimensionless conductance (in units of the conductance quantum $G_0=e^2/h$), while
$q_2/q_1=F$ delivers the Fano factor \cite{blanter:buttiker,nazarov:blanter}.

\emph{General results.---}
In order to obtain a general expression for the moment generating function,
we adopt the wave-packet variant of the scattering approach, introduced in
absence of internal dynamics in Ref.\ \cite{hassler2008}. The formalism
takes a particularly simple form if we assume that at most one electron at a
time is interacting with the internal system.
This  corresponds either to weak interactions, or to a
small attempt frequency, where the latter can be realized either by applying a small
bias voltage, or by judiciously  injecting sparse charge carriers via an
electronic pumping device \cite{turnstile}. Exploiting that under these conditions wave packets
of consecutive electrons are non-overlapping, the initial state of ${\cal N}$
electrons incident from the left lead can  be formulated as the product
of wave packets \cite{remark1}
\begin{equation}\label{eq:initial}
\Psi_{\rm initial}(\{{\bf r}_n\};t)=\left[\prod_{n=1}^{\cal N} \psi^{(\rm
in)}({\bf r}_n;t-t_n)\right]\sum_{\sigma}p_\sigma|\sigma\rangle,
\end{equation}
where $p_{\sigma}$ are the initial probability amplitudes of the internal system.
Here, $t_n\approx n \tau$ denotes the arrival times of the electrons, which
are  spread out according to the attempt frequency, and
${\cal N}\approx {\cal T}/\tau$. The final state of the reflected or transmitted electrons then
takes the form
\begin{equation}
\Psi_{\rm final}(\{{\bf r}_n\};t)=\!\!\!\!\!\sum_{\{\sigma_n\}_{n=0}^{\cal N}}\!\!
\left[p_{\sigma_0}|\sigma_{\cal N}\rangle\prod_{n=1}^{\cal N} \psi^{(\rm
out)}_{\sigma_n\sigma_{n-1}}({\bf r}_n;t-t_n)\right],\label{eq:Psifinal}
\end{equation}
where
\begin{equation}
\psi^{(\rm out)}_{\sigma'\sigma}({\bf r};t)=r_{\sigma'\sigma}({\bf r};t)+t_{\sigma'\sigma}({\bf r};t)
\end{equation}
consists of a reflected and a transmitted wave packet, confined to the left and right lead, respectively;
the indices indicate the final and initial internal state.
 These wave packets
correspond to asymptotic transmission and reflection probabilities
\begin{eqnarray}
R_{\sigma'\sigma}&=&\int d{\bf r}\, |r_{\sigma'\sigma}({\bf r}; t\gg 0)|^2,\\
T_{\sigma'\sigma}&=&\int d{\bf r}\, |t_{\sigma'\sigma}({\bf r}; t\gg 0)|^2.
\end{eqnarray}

Following the general strategy of counting statistics developed for systems
without internal degrees of freedom, we formulate the moment generating
function in terms of a generalized wave function,
$\chi(\lambda)=\langle\Psi_{\rm final}(\lambda)|\Psi_{\rm
final}(\lambda)\rangle$, where
 $\Psi_{\rm final}(\lambda)$ is of the same form as Eq.\ (\ref{eq:Psifinal}) but with $\psi^{(\rm out)}$
replaced by
\begin{equation}
\psi^{(\rm out)}_{\sigma'\sigma}({\bf r};t,\lambda)=r_{\sigma'\sigma}({\bf r};t)+e^{\lambda/2}t_{\sigma'\sigma}({\bf r};t).
\end{equation}
Here, the counting field $\lambda$ serves for book-keeping of the transmission events.
After integration over the electron coordinates,
the generating function takes the compact form of a matrix product,
\begin{equation}\chi(\lambda)= {\bf X}^T{\cal M}^{\cal N}\rho,\label{eq:chicoh}\end{equation}
where \begin{eqnarray}
{\cal M}_{\sigma'\tilde\sigma',\sigma\tilde\sigma}&=& \int d{\bf r}\, r^*_{\tilde\sigma'\tilde\sigma}({\bf r}; t)r_{\sigma'\sigma}({\bf r}; t)
\nonumber\\&&{}+e^\lambda\int d{\bf r}\, t^*_{\tilde\sigma'\tilde\sigma}({\bf r}; t)t_{\sigma'\sigma}({\bf r}; t)
\label{eq:m}
\end{eqnarray}
is the superoperator propagating the density matrix  $\rho$ of the internal system state,
generalized to include the counting field $\lambda$. Note that ${\cal M}$ can
be interpreted as a matrix where each row and column is specified by two
indices, with one index each  arising from the propagation of the bra and the
ket of the internal system state ($\rho$ then corresponds to a vector). For the pure state specified in Eq.\
(\ref{eq:initial}), the initial density matrix is
$\rho_{\sigma\tilde\sigma}=p_{\sigma}p^*_{\tilde\sigma}$;
with  modification of this initial condition, the formalism also
applies if the initial internal state is mixed. The vector ${\bf X}$ with
components
${\bf X}_{\tilde\sigma'\sigma'}=\delta_{\tilde\sigma'\sigma'}$
embodies the trace of the final density matrix. In this dynamical
description, the matrix elements $M_{\sigma',\sigma}\equiv{\cal M}_{\sigma'\sigma',\sigma\sigma}$ are
associated to the evolution of populations (the diagonal elements of $\rho$), while the other matrix elements describe corrections due to
coherence (interference of electronic wave packets generated by different
histories of the internal dynamics).

In practice, the internal degree of freedom will suffer from
decoherence due to coupling to the environment. This can be described by a
modified generalized propagator
$\widetilde{\cal M}={\cal M} {\cal Z}$,
where the diagonal matrix ${\cal Z}$ embodies suppression of off-diagonal elements
of the intermediate density matrices. If the internal system state is
continuously monitored, the internal dynamics become incoherent,
$ {\cal Z}_{\tilde\sigma'\sigma', \tilde\sigma\sigma}= \delta_{\tilde\sigma'\sigma'}
\delta_{\tilde\sigma\sigma}\delta_{\sigma\sigma'}$,
and the time evolution reduces to a classical stochastic problem described by
a rate equation, which only involves transition probabilities between
populations. The generating function then simplifies to
\begin{equation}\chi(\lambda)= {\bf Y}^TM^{\cal N}{\bf P},\label{fullgeneratingfunction}\end{equation}
where
\begin{eqnarray}
M_{\sigma'\sigma}=R_{\sigma'\sigma}+e^\lambda T_{\sigma'\sigma}
\label{eq:m2}
\end{eqnarray}
is the classical propagator of the population probabilities with initial conditions
${\bf P}_\sigma=|p_\sigma|^2$,
while the vector  ${\bf Y}^T=(1,1,1,\ldots)$ with unit components
embodies the summation over all final states of the system \cite{remark2}.

Of particular interest is the steady state behaviour of the system,
corresponding to ${\cal T}\gg \tau$. To analyze this limit we
diagonalize the matrices ${\cal M}$ (for coherent internal dynamics),  $M$
(for incoherent internal dynamics), or  $\widetilde {\cal M}$ (for cases in
between these two limiting scenarios), and denote the largest eigenvalue by
$D_{\rm max}(\lambda)$. For large times, the cumulant generating function
then becomes
\begin{equation}\label{eq:chieval}
\ln \chi(\lambda)\sim \frac{{\cal T}}{\tau}\ln  D_{\rm max}(\lambda).
\end{equation}
Therefore, the problem of full counting statistics is reduced to finding the eigenvalues of
the matrices  ${\cal M}$, $M$, or $\widetilde {\cal
M}$, respectively. Expanding $D_{\rm max}(\lambda)=\sum_{k=0}^\infty
d_k \lambda^k/k!$ as a Taylor series, the dimensionless conductance is given
by $g=d_1$, and the Fano factor is given by $F=(d_2-d_1^2)/d_1$  \cite{remark3}.

\emph{Dynamical electron bunching.---}
In the conventional case of phase-coherent electronic transport through
passive devices, shot noise is suppressed due to the Pauli exclusion
principle, resulting in a Fano factor $F<1$ which indicates sub-Poissonian
noise statistics \cite{blanter:buttiker,nazarov:blanter}. As we now show, even though we consider electrons that
never directly interact with each other, the internal system degrees of
freedom can induce dynamical correlations which result in super-Poissonian
noise with $F\gg 1$, corresponding to dynamical bunching of electrons. This
possibility is most transparently revealed in the case of incoherent dynamics
of an internal two-state system. In anticipation of the concrete example
discussed below, we denote the two internal states as $\sigma=+,-$. The matrix $M$ then takes the form
\begin{equation}M=\left(\begin{array}{cc}
R_{++}+e^{\lambda}T_{++} &R_{+-}+e^{\lambda}T_{+-} \\
 R_{-+}+e^{\lambda}T_{-+}  &R_{--}+e^{\lambda}T_{--}    \end{array} \right),
\end{equation}
and the largest eigenvalue can be calculated explicitly.
Consider now the case of small switching probabilities
$S_{+-}=R_{+-}+T_{+-}\ll 1$, $S_{-+}=R_{-+}+T_{-+}\ll 1$. On average, the two
states are then populated with probabilities $P_+=S_{+-}/(S_{+-}+S_{-+})$ and
$P_-=S_{-+}/(S_{+-}+S_{-+})$. In terms of these probabilities, the
dimensionless conductance takes the simple form $g\approx P_+T_{++}+P_-T_{--}$ of a
time-averaged transmission probability. Assuming $T_{++}\neq T_{--}$, the
Fano factor \begin{equation}
F\approx\frac{1}{S_{+-}+S_{-+}}\frac{2 P_+ P_-
(T_{++}-T_{--})^2}{P_+T_{++}+P_-T_{--}},
\label{fanomax}
\end{equation}
 on the other hand, can become
arbitrarily large. The noise is then dominated by the rare probabilistic
switching between states of different transmission probabilities. In the
limiting case that one of the two transmission probabilities is much larger
than the other (say,  $T_{++} \gg T_{--}$), the device is partially
transmitting during some time intervals, but  effectively shut down during
others,  which results in well spaced-out trains of transmitted electrons.
These observations also extend to multi-state dynamics; noise is generally
enhanced whenever the transmission probabilities differ between the states of the device.

\emph{Application to a mesoscopic device.---}
We now illustrate how conditions of dynamically enhanced current fluctuations
can be realized in a specific mesoscopic device, the coherent which-path interferometer shown in Fig.\ \ref{fig:1}(b)  \cite{schomerus:noat,schomerus:robinson}. The internal degree of freedom
arises from an excess electron in a double quantum dot, with states $|\uparrow\rangle$, $|\downarrow\rangle$ for occupation of the upper or lower dot, respectively. Depending on its location, this electron blocks the path of charge carriers moving through the upper or lower arm of an Aharonov-Bohm ring, where coherent transport can be tuned by varying the magnetic flux $\Phi$.
We assume that the isolated system eigenstates are the symmetric and antisymmetric orbital states $|\pm\rangle=(|\uparrow\rangle\pm|\downarrow\rangle)/\sqrt{2}$ of the excess electron, and that they are separated by a tunnel splitting energy $\Delta$ (corresponding to ${\cal E}_\pm=\mp \Delta/2$).

To characterize the transport through this device, we calculate the matrix elements (\ref{eq:m}) by numerical propagation of initially Gaussian wave packets.
The conductance and Fano factor are then obtained from
the eigenvalues of the matrices ${\cal M}(\lambda)$  (for coherent internal dynamics) and  $M(\lambda)$   (for incoherent internal dynamics), as described above.
Figure \ref{fig:results}(a) shows the resulting Fano factor as a function of  $\Phi$ for a set of system parameters.
We observe that the current fluctuations can be significantly enhanced  above the threshold $F=1$ for super-Poissonian noise.
For $\Phi\to 0$, the switching probabilities $S_{+-}$,  $S_{-+}$ vanish because of the parity symmetry of the system \cite{schomerus:robinson}, and the Fano factor increases over all bounds, as predicted by Eq.\ (\ref{fanomax}).
In this regime, the Fano factor discriminates coherent from incoherent dynamics (bottom inset; as shown in the top inset, the conductance is unaffected by this distinction).
In contrast, when half a flux quantum penetrates the ring ($\Phi=h/2e$),  parity symmetry  ensures  that
each transmission event is strictly correlated with a transition of the internal system degree of freedom, while no such transitions occur when an electron is reflected (i.e., $R_{-+}=R_{+-}=T_{++}=T_{--}=0$) \cite{schomerus:robinson}. Even in this case, dynamical electron bunching can occur: The maximal Fano factor for incoherent internal dynamics now is  $F=2$, which is realized when
$R_{--},T_{-+}\ll R_{++},T_{+-}$ (or vice versa), so that $P_+\gg P_-$. The transport then consists of pairs of transmitted electrons, which coincide with the rare transitions into the state $|-\rangle$ (first transmitted electron), followed by its immediate depopulation (second transmitted electron).
Figure \ref{fig:results}(b) shows that such conditions can be met by tuning the tunnel splitting $\Delta$, and is insensitive
to the degree of coherence in the internal dynamics.
Guided by these results, we expect that super-Poissonian noise statistics should be common when mesoscopic systems possess internal degrees of freedom.

\emph{Concluding remarks.---}In summary, we have investigated the phase coherent transport through mesoscopic devices with internal degrees of freedom, and formulated the full counting statistics in terms of a generalized propagator of the density matrix for these freedoms [Equations (\ref{eq:chicoh})--(\ref{eq:chieval})].
In general, the counting statistics depend on the degree of coherence of the internal system dynamics.
The dynamical switching of the internal system degrees of freedom
can  enhance the current fluctuations, resulting in Fano factors that significantly exceed the threshold $F=1$ for super-Poissonian noise.


\begin{thebibliography}{99}
\bibitem{remark1} Because wave packets are
non-overlapping, the  analysis does not require to anti-symmetrize
the wave function.
\bibitem{remark2} The reduction from ${\cal M}$ to $M$ also occurs when inelastically scattered wave packets do not energetically overlap.
\bibitem{remark3}
The leading
coefficient in this expansion is $d_0=1$, as follows from the normalization
of the wave function $\Psi$. This eigenvalue corresponds to the left
eigenvector ${\bf X}$ of the matrices ${\cal M}$ and $\widetilde{\cal
M}$, or ${\bf Y}$ for the matrix $M$,  at $\lambda=0$.
\end{thebibliography}
\end{document}